\documentclass[prl,a4paper,twocolumn,showpacs]{revtex4}

\newcommand{\nc}{\newcommand}

\nc{\be}[1]{\begin{equation}\mbox{$\label{#1}$}}
\nc{\bea}[1]{\begin{eqnarray} \mbox{$\label{#1}$}}
\nc{\Section}[2]{\section{#2}\label{#1}}
\nc{\Bibitem}[1]{\bibitem{#1}}
\nc{\Label}[1]{\label{#1}}

\nc{\eea}{\end{eqnarray}}
\nc{\ee}{\end{equation}}

\nc{\bdm}{\begin{displaymath}}
\nc{\edm}{\end{displaymath}}
\nc{\dpsty}{\displaystyle}
\nc{\bc}{\begin{center}}
\nc{\ec}{\end{center}}
\nc{\ba}{\begin{array}}
\nc{\ea}{\end{array}}
\nc{\bab}{\begin{abstract}}
\nc{\eab}{\end{abstract}}
\nc{\btab}{\begin{tabular}}
\nc{\etab}{\end{tabular}}
\nc{\bit}{\begin{itemize}}
\nc{\eit}{\end{itemize}}
\nc{\ben}{\begin{enumerate}}
\nc{\een}{\end{enumerate}}
\nc{\bfig}{\begin{figure}}
\nc{\efig}{\end{figure}}

\nc{\arreq}{&\!=\!&}
\nc{\arrmi}{&\!-\!&}
\nc{\arrpl}{&\!+\!&}
\nc{\arrap}{&\!\!\!\approx\!\!\!&}
\nc{\non}{\nonumber}
\nc{\align}{\!\!\!\!\!\!\!\!&&}

\def\lsim{\; \raise0.3ex\hbox{$<$\kern-0.75em
      \raise-1.1ex\hbox{$\sim$}}\; }
\def\gsim{\; \raise0.3ex\hbox{$>$\kern-0.75em
      \raise-1.1ex\hbox{$\sim$}}\; }

\nc{\DOT}{\hspace{-0.08in}{\bf .}\hspace{0.1in}}
\nc{\Laada}{\hbox {$\sqcap$ \kern -1em $\sqcup$}}
\nc\loota{{\scriptstyle\sqcap\kern-0.55em\hbox{$\scriptstyle\sqcup$}}}
\nc\Loota{{\sqcap\kern-0.65em\hbox{$\sqcup$}}}
\nc\laada{\Loota}
\nc{\qed}{\hskip 3em \hbox{\BOX} \vskip 2ex}

\nc{\real}{{\rm I \! R}}
\nc{\Z}{{\sf Z \!\!\! Z}}
\nc{\complex}{{\rm C\!\!\! {\sf I}\,\,}}
\def\bigid{\leavevmode\hbox{\small1\kern-3.8pt\normalsize1}}
\def\id{\leavevmode\hbox{\small1\kern-3.3pt\normalsize1}}
\nc{\slask}{\!\!\!/}
\nc{\bis}{{\prime\prime}}
\nc{\pa}{\partial}
\nc{\na}{\nabla}
\nc{\ra}{\rangle}
\nc{\la}{\langle}
\nc{\goto}{\rightarrow}
\nc{\swap}{\leftrightarrow}

\nc{\EE}[1]{ \mbox{$\cdot10^{#1}$} }
\nc{\abs}[1]{\left|#1\right|}
\nc{\at}[2]{\left.#1\right|_{#2}}
\nc{\norm}[1]{\|#1\|}
\nc{\abscut}[2]{\Abs{#1}_{\scriptscriptstyle#2}}
\nc{\vek}[1]{{\rm\bf #1}}
\nc{\integral}[2]{\int\limits_{#1}^{#2}}
\nc{\inv}[1]{\frac{1}{#1}}
\nc{\dd}[2]{{{\partial #1}\over{\partial #2}}}
\nc{\ddd}[2]{{{{\partial}^2 #1}\over{\partial {#2}^2}}}
\nc{\dddd}[3]{{{{\partial}^2 #1}\over
    {\partial #2 \partial #3}}}
\nc{\dder}[2]{{{d #1}\over{d #2}}}
\nc{\ddder}[2]{{{d^2 #1}\over{d {#2}^2}}}
\nc{\dddder}[3]{{d^2 #1}\over
    {d #2 d #3}}
\nc{\dx}[1]{d\,^{#1}x}
\nc{\dy}[1]{d\,^{#1}y}
\nc{\dz}[1]{d\,^{#1}z}
\nc{\dl}[1]{\frac{d\,^{#1}l}{(2\pi)^{#1}}}
\nc{\dk}[1]{\frac{d\,^{#1}k}{(2\pi)^{#1}}}
\nc{\dq}[1]{\frac{d\,^{#1}q}{(2\pi)^{#1}}}

\nc{\bfT}{{\bf T }}

\nc{\cA}{{\cal A}}
\nc{\cB}{{\cal B}}
\nc{\cD}{{\cal D}}
\nc{\cE}{{\cal E}}
\nc{\cG}{{\cal G}}
\nc{\cH}{{\cal H}}
\nc{\cL}{{\cal L}}
\nc{\cO}{{\cal O}}
\nc{\cT}{{\cal T}}
\nc{\cN}{{\cal N}}
\nc{\cR}{{\cal R}}
\nc{\rvac}[1]{|{\cal O}#1\rangle}
\nc{\lvac}[1]{\langle{\cal O}#1|}
\nc{\rvacb}[1]{|{\cal O}_\beta #1\rangle}
\nc{\lvacb}[1]{\langle{\cal O}_\beta #1 |}
\nc{\bb}{\bar{\beta}}
\nc{\bt}{\tilde{\beta}}
\nc{\ctH}{\tilde{\cal H}}
\nc{\chH}{\hat{\cal H}}
\nc{\1}{\aa}
\nc{\2}{\"{a}}
\nc{\3}{\"{o}}
\nc{\4}{\AA}
\nc{\5}{\"{A}}
\nc{\6}{\"{O}}
\nc{\al}{\alpha}
\nc{\g}{\gamma}
\nc{\Del}{\Delta}
\nc{\e}{\textrm{e}}
\nc{\eps}{\epsilon}
\nc{\lam}{\lambda}
\nc{\Om}{\Omega}
\nc{\ve}{\varepsilon}
\nc{\mn}{{\mu\nu}}
\nc{\vp}{\varphi}

\nc{\rf}[1]{(\ref{#1})}
\nc{\nn}{\nonumber \\*}
\nc{\bfB}{\bf{B}}
\nc{\bfv}{\bf{v}}
\nc{\bfx}{\bf{x}}
\nc{\bfy}{\bf{y}}
\nc{\vx}{\vec{x}}
\nc{\vy}{\vec{y}}
\nc{\oB}{\overline{B}}
\nc{\oI}{\overline{I}}
\nc{\oR}{\overline{R}}
\nc{\rar}{\rightarrow}
\nc{\ti}{\times}
\nc{\slsh}{\hskip-5pt/}
\nc{\sm}{Standard~Model~}
\nc{\MP}{M_{\rm Pl}}
\nc{\mpl}{M_{\rm Pl}}
\nc{\tp}{t_{\rm Pl}}

\nc{\pmin}{p_{\rm min}}
\nc{\pmax}{p_{\rm max}}
\nc{\fo}{f_0}
\nc{\foi}{f_{0,i}\,}
\nc{\fop}{f_0^P}
\nc{\fou}{f_0^U}

\nc{\eff}{{\rm eff}}
\nc{\MT}{M_{\rm T}}
\nc{\ML}{M_{\rm L}}
\nc{\kk}{\vek{k}}
\nc{\pp}{{\rm p}}
\nc{\pt}{\partial_t}
\nc{\half}{{1\over 2}}
\nc{\w}{\omega}
\nc{\uhat}{\hat{U}_\w}

\nc{\etal}{\mbox{\it et al.}}
\nc{\ie}{{\it i.e. }}
\nc{\eg}{{\it e.g. }}
\nc{\trh}{T_{\rm RH}}
\nc{\ad}{{a'\over a}}
\nc{\bd}{{b'\over b}}
\nc{\Rd}{{R'\over R}}
\nc{\diag}{{\textrm{diag}}}
\nc{\mato}[1]{\tilde{#1}}
\nc{\sech}{\textrm{sech}}
\nc{\I}{\textrm{I}}
\nc{\II}{\textrm{II}}
\nc{\III}{\textrm{III}}
\nc{\vev}[1]{\langle #1 \rangle}
\nc{\hyp}{\,\; F_{1{\hskip -16pt}2}{\hskip 11pt}}
\nc{\brhom}{\overline{\rho}_M}
\nc{\brho}{\overline{\rho}}
\nc{\rhob}{\overline{\rho}}
\nc{\Pb}{\overline{P}}
\nc{\bH}{\overline{H}}
\nc{\ep}{{1+4\eps}}

\nc{\lcdm}{$\Lambda$CDM}
\nc{\omm}{\Omega_m}

\def\smiley{\hbox{\large$\bigcirc$\hspace{-.80em}%
\raise.2ex\hbox{$\cdot\cdot$}\kern-.61em    
\lower.2ex\hbox{\scriptsize$\smile$}}\ }

\def\frowney{\hbox{\large$\bigcirc$\hspace{-.80em}%
\raise.2ex\hbox{$\cdot\cdot$}\kern-.635em
\lower.2ex\hbox{\scriptsize$\frown$}}\ }

\begin{document}

\title{Cosmological expansion and the uniqueness of gravitational action}

\author{T. Multam\"aki}
\thanks{tuomas@nordita.dk}
\affiliation{NORDITA, Blegdamsvej 17, DK-2100 Copenhagen, DENMARK}
\author{I. Vilja}
\thanks{vilja@utu.fi}
\affiliation{Department of Physics,
University of Turku, FIN-20014 Turku, FINLAND}

\pacs{04.50.+h, 98.80.-k}
\preprint{NORDITA-2005-43}
\begin{abstract}

Modified theories of gravity have recently been studied by several
authors as possibly viable alternatives to the cosmological
concordance model. Such theories attempt to explain the accelerating 
expansion of the universe
by changing the theory of gravity, instead of introducing dark energy. 
In particular, a class of models based on higher order curvature
invariants, so-called $f(R)$ gravity models, has drawn attention. 
In this letter we show 
that within this framework, the expansion history of the universe does not 
uniquely determine the form of the gravitational action and it can be radically
different from the standard Einstein-Hilbert action. 
We demonstrate that for any barotropic fluid, there
always exists a class of $f(R)$ models that will have
exactly the same expansion history as that arising from the 
Einstein-Hilbert action. 
We explicitly show how one can extend the Einstein-Hilbert action 
by constructing a $f(R)$ theory that is equivalent on the classical level. 
Due to the classical equivalence
between $f(R)$ theories and Einstein-Hilbert gravity with an extra scalar field, 
one can also hence construct equivalent scalar-tensor theories with standard expansion.

\end{abstract}
\maketitle


\section{Introduction}

One of the great surprises in modern cosmology has been the
accelerating expansion of the universe. The first sign of 
this unexpected behaviour came from distant supernova observations (SNIa)
in 1998, when it was discovered that far away supernovae are dimmer than
expected \cite{snia}. The SNIa observations have since improved,
notably by a number very distant supernova observations by the Hubble
Space Telescope, and the main conclusion still stands: a critical density 
matter dominated universe is not consistent with the data.
Such a universe, also known as the Einstein-deSitter (EdS) model, 
was also disfavoured by independent observations
based on measuring the small fluctuations in the cosmic
microwave background (CMB) \cite{cmb}. Combining these two
observations along with a third independent observation based
on the clustering of galaxies on very large scales \cite{lss},
the present day cosmological model emerges.
The cosmological concordance model is a geometrically flat \lcdm\ model,
which is a critical density universe dominated by cold dark matter 
and dark energy in the form of a cosmological constant. 
Furthermore, cross-correlating the observations of galaxy clusters 
with CMB measurements,
one has seen evidence of the decay of gravitational potentials
on very large scales \cite{isw}, as predicted in the \lcdm\ model.
Although independent measurements are statistically not very significant, 
combining measurements leads to a picture that is very well consistent
with the concordance \lcdm\ model \cite{iswcomb}.

A natural alternative to adding a mysterious new form of energy to
our physical picture of the universe, is to consider modifications
of general relativity. This is also motivated by the fact that 
we only have precision observations of gravity from sub-millimeter
scales to solar system scales, which is very far from the 
present Hubble radius that is the scale relevant to the
question of dark energy.

A particular class of models that has drawn a significant amount
attention recently are the so-called $f(R)$ gravity models
(see \eg \cite{turner,turner2,allemandi,meng,nojiri3} and references therein).
They form a class of higher derivative gravity theories that
include higher order curvature invariants in the gravitational
action. Such an extension of the Einstein-Hilbert (EH) Lagrangian can
be viewed natural as there is no {\it a priori} reason why the gravitational
action should be linear in the Ricci scalar $R$. Furthermore,
higher order terms can naturally appear in low energy effective
Lagrangians of quantum gravity and string theory.

After $f(R)$ theories were proposed as an alternative solution 
to the dark energy problem \cite{cappo1}-\cite{nojiri2}, it was quickly
realized that two major obstacles exist. Firstly, some of these
theories exhibit instabilities that will render space-time
unstable \cite{dolgov}. 
A second serious but more practical
hindrance has been the difficulty in actually integrating the
equations of motion numerically and comparing to observational
data. 

Another twist in the tale is the question of frames and conformal
transformations. While the Einstein and Jordan frames are mathematically 
equivalent on the classical level, the physical equivalence of the two 
frames at the perturbation and quantum level has been discussed in 
reference to $f(R)$ theories. Possible restrictions arise due to 
non-standard gravitational effects  constrained by fifth force 
experiments (see \cite{confprobs} and references therein).

Recently, an inverse approach to the problem of integrating the complicated
equations of motion arising from a general $f(R)$ action
has been proposed in \cite{cappo3}. Instead of specifying a particular form 
of $f(R)$ one considers the inverse problem of reconstructing $f(R)$ from
the expansion history of the universe. 

In this letter we are also concerned with the inverse problem of
reconstructing $f(R)$, given an expansion history.
We show that for any barotropic equation state, the functional
form of $f(R)$ and hence the gravitational action, is not
uniquely determined by the expansion history of the universe.
Instead, for a given fluid one can always construct a class of
gravitational actions that will have the same cosmological
evolution as the EH action. As an example, we 
demonstrate this explicitly for cosmologically relevant solutions.


\section{Basic formalism of $f(R)$ gravity}

The action for $f(R)$ gravity is (with $8\pi G=1$) (see \eg \cite{cappo3})
\be{action}
S = \int{d^4x \sqrt{-g} \Big( f(R) + {\cal{L}}_{m} \Big)},
\ee
form which it follows in the standard metric approach (as opposed to the
so-called Palatini approach)
\be{eequs}
G_{\mu\nu}=R_{\mu\nu}-\frac 12 R g_{\mu\nu}=T^c_{\mu\nu}+T^m_{\mu\nu},
\ee
where
\bea{tmunu}
T^c_{\mu\nu} & = & \frac{1}{f'(R)}\Big\{\frac 12 g_{\mu\nu}\Big(f(R) - R f'(R)
\Big)+\nonumber \\
& + & f'(R)^{;\mu\nu}\Big(g_{\alpha\mu}g_{\beta\nu}-g_{\mu\nu}
g_{\mu\nu}\Big)\Big\}).
\eea
The standard minimally coupled stress-energy tensor, 
$\tilde{T}_{\mu\nu}^m$, from the matter Lagrangian, $\mathcal{L}_m$ 
in the action, Eq. (\ref{action}), is related to $T_{\mu\nu}^m$ by
$T^m_{\mu\nu}=\tilde{T}^m_{\mu\nu}/f'(R)$.
The equations of motion arising from the action in a 
Friedmann-Robertson-Walker universe are
\be{fried1}
H^2 +\frac{k}{a^2}= \frac{1}{3} \Big(\rho_{c} + \frac{\rho_m}{f'(R)} \Big)
\ee
and
\be{fried2}
2 \dot{H}+3 H^2 +\frac{k}{a^2}=  - \left ( p_{c} + \frac{p_m}{f'(R)} \right ),
\ee
where the energy density and pressure of the curvature fluid are
\bea{curfluid}
\rho_{c} & = & \frac{1}{f'(R)} \Big \{ \frac{1}{2} \Big(f(R)-R f'(R)\Big)-3H \dot{R} f''(R) \Big\}\nonumber\\
p_{c} & = & {1\over f'(R)}\Big\{\dot{R}^2f'''(R)+2H\dot{R}f''(R)+\ddot{R}f''(R)\nonumber\\
& & \ \ \ \ \ \ \ \ \ \ -\frac 12 \Big(f(R)-R f'(R)\Big)\Big\}.
\eea
In addition to these, we also have a constraint equation
for the curvature scalar
\be{const}
f''(R) \Big\{ R + 6 \Big(\dot{H} +2H^2 +\frac{k}{a^2}\Big)\Big\} = 0
\ee
and we can write the continuity equation of the total fluid 
from Eqs.\ (\ref{fried1}) and (\ref{fried2}) as
$\dot{\rho}+3H(\rho+p)=0,$
where $\rho=\rho_c+\rho_m/f'(R)$ and $p=p_c+p_m/f'(R)$.

From the Einstein's equations we see that due to the presence of the
$\ddot{R}$ term in $p_c$, we have a fourth order differential equation 
system for the scale factor of $a(t)$. Given a gravitational theory, or given
the form of $f(R)$, one could hope to solve the arising differential equation.
However, even for simple choices of $f(R)$, such as $f(R)=R+\mu^4/R$, the resulting
differential equation (see eg.\ \cite{turner}) is complicated and one must 
resort to numerical methods. The numerical solutions are also problematic
since one then needs information about the higher derivatives of the
scale factor. 


\section{Uniqueness of the gravitational action}

It is clear that for $f(R)=R$, Equations (\ref{fried1}) and (\ref{fried2})
reduce to the standard Friedmann equations of Einstein gravity,
since then $\rho_{c}=p_{c}=0$. However, this does not guarantee 
that $f(R)=R$ is the only choice that reduces to the standard Friedmann 
equations but in general standard equations are reached if conditions
\bea{stcond}
\rho_c+\frac{\rho_m}{f'(R)} & = & \rho_m,\label{1cond}\\
p_c+\frac{p_m}{f'(R)} & = & p_m
\eea
hold.
Assuming that the continuity equation holds independently for 
the ordinary matter fluid, $\dot{\rho}_m+3H(\rho_m+p_m)=0$, 
these two conditions are in fact equivalent.

The fact that one can recover the standard Friedmann equations
is a consequence of higher derivatives in the Einstein's equations.
The general Friedman equations are higher order differential 
equations than the standard equations. Thus the standard solutions can also
be solutions of the general equations without violating the uniqueness of
the solutions of differential equations.
One can step even a little further and show the standard Friedmann
equations supplemented with Eq.\ (\ref{stcond}) are enough that  
both general Eqs.\ (\ref{fried1}) and (\ref{fried2}) are fulfilled.

When the ordinary Einstein's equations hold, the curvature scalar can
be written as $R=-\rho_m+3p_m$, so that for any barotropic
equation of state (expect for the radiation dominated case),
$p_m=p_m(\rho_m)$, the curvature scalar
is a function of the matter density, $R=R(\rho_m)$.
One can then write, at least formally, $\rho_m=\rho_m(R)$, \eg for
cold dark matter $\rho_m=-R$. Therefore, the quantity
$H\dot{R}$ appearing in the expression for $\rho_c$
can be written as 
$H\dot{R}= -(\rho_m-3k\, a^{-2})(\rho_m+p_m)R'(\rho_m)$.
In a flat universe, the curvature fluid is hence
a function of $R$ only, $\rho_c=\rho_c(R)$ and 
therefore the condition, Eq.\ (\ref{1cond}), 
gives a second order homogeneous differential equation for 
$g(R)\equiv -R+f(R)$:
\be{genflat}
3\frac{\rho_m}{\rho'_m}\Big(\rho_m+p_m\Big)g''(R)-
\Big(\frac 12 R+\rho_m\Big)g'(R)+\frac 12 g(R)=0,
\ee
where $\rho_m$ and $p_m$ are given as functions of $R$ only.
Hence
for any barotropic fluid, there exists a general $f(R)$
with two integration constants (expect if $\rho_m+p_m=0$ which
leads to a first order differential equation) that will lead to the exactly 
same background evolution as in the standard EH gravity. 
In other words, for any barotropic equation of state $f(R)$ 
gravity constructed in this way,
will always have among its solutions the same solution as the 
solution arising from EH gravity.

In a non-flat universe, the conclusions are unchanged but 
the technical details are somewhat more complicated since now  
one cannot write $H\dot{R}$ as a function of $R$ only 
due to the presence of the $k$-term. Instead, one needs
to express everything in terms of the scale factor $a$, resulting
in a differential equation for $F(a)=f(R(a))$ to be solved. 
After inverting $R=R(a)$
one obtains an expression for $f(R)$.


\section{Constructing equivalent gravitational actions}

To illustrate the general conclusions of the previous section,
we  construct explicit gravitational
actions that have identical background expansion as the
solution arising from the EH action among their solutions. 

Consider first a fluid with a constant equation of state, $p_m=w\rho_m$, 
where $w$ is a arbitrary constant, excluding cases $w=-1,\, 1/3$. 
Now $\rho_m=\rho_{m,0}a^{-3(1+w)}$, 
and as explained above, expressing 
all in terms of $R$, writing $g(R)=G(u)$, 
where $u=R/(\rho_{m,0}(3w-1))$, we obtain the general equation for $G$:
\bea{ge}
0&=& 6\,{w+1\over 3w-1}
\left( 1 - \tilde k\, u^{- {3w+1\over 3(w+1)}}\right) u^2\, G''(u)\nonumber\\
& & - {3w+1\over 3w-1} u\, G'(u) + G(u),
\eea 
where $\tilde k= k/\rho_{m,0}$. 
In the case of a flat universe, $k=0$, Eq. (\ref{ge}) can be
solved:
\be{sol}
g(R)=c_+ (-R)^{\alpha_+(w)} +c_- (-R)^{\alpha_-(w)},
\ee
where
$\alpha_\pm(w)=(9w +7 \pm \sqrt{9w^2 +78 w +73})/(12(w+1))$
and $c_\pm$ are constants (note the $R$ is negative with our conventions). 
In contrast to the $k\ne 0$ case, the spatially flat solution does not depend on the 
(present day) density $\rho_{m,0}$. In the special case of a flat matter 
dominated universe, $w=0$, $k=0$, the solution translates to 
\bea{finalf}
f(R)=R & + & c_+ (-R)^{(7+\sqrt{73})/12}\nonumber\\
& + & c_- (-R)^{(7-\sqrt{73})/12}.
\eea
Note how the other solution actually grows with $|R|$ \ie its effect 
is larger at early times.

The radiation dominated case $w=1/3$ is not included to the solution 
given above because now $R\equiv 0$ and the scale parameter can not be 
expressed in terms of $R$.  However, this means that for radiation 
domination $\dot R \equiv 0$ and therefore from Eqs (\ref{stcond}) 
and (\ref{curfluid}) any function $g(R)$ with 
$g(0)=g'(0)=0$ will give the same radiation dominated expansion. 
The other excluded value $w=-1$, corresponding to the deSitter model,
gives $R=-4\rho_\Lambda$. Hence $\dot R =0$ and Eq. (\ref{genflat}) 
simplifies to $R\, g'(R)-2 g(R)=0$ giving $g(R)= c R^2$. 
This solution also coincides with the limiting 
value of $\alpha_+$ -solution: $\alpha_+(w\goto -1) \goto 2$.

Another interesting example is the flat \lcdm\ model
\be{lcdm}
a(t)=\left(\frac{\omm}{\Omega_\Lambda}\right)^{1/3}\sinh^{2/3}
(\frac 32 \sqrt{\Omega_\Lambda} H_0 t),
\ee
in which case the differential equation for $g(R)$ reads
as
\be{lcdmequ}
(R+3\beta)(R+4\beta)g''(R)-(\frac 16 R+\beta)g'(R)-\frac 16 g(R)=0,
\ee
where we have defined $\beta\equiv 3\Om_\Lambda H_0^2$.
This equation has a solution in terms of the hypergeometric
function $F(\alpha,\beta ;\gamma ; z)$ and the solution in terms
of $f(R)$ is
\bea{lcdmsol}
f(R) & = & R+c_+ z^{\alpha_+}
F(-\alpha_+, \frac 32 - \alpha_+;\alpha_--\alpha_++1;\frac 1z)\nonumber\\ 
&+&c_- z^{\alpha_-}
F(-\alpha_-, \frac 32 - \alpha_-;\alpha_+-\alpha_-+1;\frac 1z)
\eea
where $z\equiv -3-R/\beta$ and $\alpha_\pm\equiv \alpha_\pm(w=0)$.
It is easy to see that in the limit $R\goto -\infty$ (or $z\goto +\infty$)
we recover the matter dominated case (\ref{finalf}) as expected.


\section{Stability}

An important question in considering $f(R)$ models as realistic theories
of gravity is the stability of the ground state \cite{dolgov}.
The stability criterion can be formulated as a condition
on the sign of the potential given by \cite{nojiri}
\bea{pot}
U(R)&=& \frac 13 R - {f^{(1)}(R)f^{(3)}(R) R \over 3 f^{(2)}(R)^2}
-{f^{(1)}(R) R\over 3 f^{(2)}(R)},\nonumber\\
&-& {f^{(1)}(R)\over 3 f^{(2)}(R)}+{2 f(R)f^{(3)}(R)\over 3 f^{(2)}(R)^2}
\eea
where we have assumed that the universe is homogeneous (note that 
we have different metric conventions compared to \cite{nojiri}). 
If $U(R)>0$ for
classical, unperturbed solution, then the linear perturbations $\delta R$ 
are oscillatory without exponentially decreasing or decaying modes. 
In terms of function $g(R)$ the stability condition obtained from 
Eq. (\ref{pot}) reads as (for $g''\neq 0$)
\be{pot2}
\Big(2\, g-R\,g'\Big)g^{(3)}
-\Big(1+g'-R\, g''\Big)g''
>0.
\ee
Equation  (\ref{pot2}) shows that for the radiation dominated case 
gravitational stability  is achieved simply if $g''(0)<0$ and 
for the deSitter case (where $g=c\, R^2$) if $c<0$. 
In order to have an understanding whether stable choices of $f(R)$
exist in the more general case where $p=w\rho$, we consider
the two solutions
$f_{\pm}(R)=R+c_\pm (-R)^{\alpha_\pm}$ separately and leave more
general considerations for future work.
We find that both solutions are stable in two separate regions,
\bea{stab}
c_\pm&<&0,\nonumber\\
 c_\pm &>& 
{6(1+w)(-R)^{1-\alpha_\pm(w)} \over 17+15 w \mp \sqrt{ 9 w^2+ 78 w + 73}}
\eea
and hence one can always construct stable extensions
of the EH action for such a fluid.


\section{Conclusions and discussion}

In the present letter we have considered the uniqueness of
the gravitational action among $f(R)$ gravity models.
We have showed that for any barotropic fluid, the evolution 
of the scale factor of the universe does not uniquely
determine the form of $f(R)$, and hence nor the gravitational action.
Instead, one can always construct a $f(R)$ theory that will
have the same background evolution among its solutions.
In other words, cosmology covered by standard Friedmann equations has always
a generalized counterpart with the very same classical evolution independently
of the actual equation of state of the ordinary matter. 

As an example, we have constructed explicit forms of $f(R)$ in a
number cosmologically interesting cases. By considering small
fluctuations around the ground state, we have shown that
such modified theories of gravity are also stable.

In this letter we have not directly addressed the dark energy problem.
However, our results are likely to  be relevant to such considerations.
Assuming that we have a matter dominated universe, we now know
the form of $f(R)$ we need in order to reproduce the matter dominated
regime exactly. This will guide us in considering choices of $f(R)$
that will lead to late time acceleration while preserving standard
matter dominated expansion at early times.

These considerations are also closely related to the models
obtained by conformal transformations. Although the gravitational
actions constructed in this article may be rather unintuitive,
their counterparts in scalar-tensor -gravity may seem natural. Moreover  
from the analysis presented here we know that some highly non trivial,
probably non-minimally coupled  scalar-tensor theories reproduce exactly
the standard evolution of the background metric.

Since the background expansion alone cannot distinguish between
different choices of $f(R)$, one must study perturbations in such cosmologies.
This will ultimately guide us in choosing the correct gravitational
action which is possibly not the simple Einstein-Hilbert action.




\begin{thebibliography}{X}


\bibitem{snia} A.~G.~Riess {\it et al.}, 
Astron.\ J.\  {\bf 116}, 1009 (1998);
S.~Perlmutter {\it et al.}, Astrophys.\ J.\  {\bf 517}, 565 (1999).

\bibitem{cmb} D.~N.~Spergel {\it et al.} , Astrophys.\ J.\ Suppl.\  
{\bf 148}, 175 (2003).  

\bibitem{lss} M.~Tegmark {\it et al.}, 
Phys.\ Rev.\ D {\bf 69}, 103501 (2004);
G.~Efstathiou {\it et al.}, Mon.\ Not.\ Roy.\ Astron.\ Soc.\  {\bf 330}, 
L29 (2002).  

\bibitem{isw} S.~Boughn and R.~Crittenden, Nature, {\bf 472},
45 (2004); M.~R.~Nolta {\it et al.}, Astrophys.\ J.\  {\bf 608}, 10 (2004);
P.~Fosalba and E.~Gazta\~naga, Mon.\ Not.\ Roy.\ Astron.\ Soc.\  {\bf 350}, L37
(2004);
P.~Fosalba, E.~Gazta\~naga and F.~Castander, Astrophys.\ J.\  {\bf 597}, 
L89 (2003);
N.~Afshordi, Y.~S.~Loh and M.~A.~Strauss, Phys.\ Rev.\ D {\bf 69}, 
083524 (2004);
R.~Scranton {\it et al.}, arXiv:astro-ph/0307335.

\bibitem{iswcomb}  E.~Gazta\~naga, M.~Manera and T.~Multam\"aki, 
arXiv:astro-ph/0407022;  
P.~S.~Corasaniti, T.~Giannantonio and A.~Melchiorri, 
Phys.\ Rev.\ D {\bf 71}, 123521 (2005).

\bibitem{turner}  S.~M.~Carroll, V.~Duvvuri, M.~Trodden and M.~S.~Turner,  
Phys.\ Rev.\ D {\bf 70}, 043528 (2004).  

\bibitem{turner2}  S.~M.~Carroll, A.~De Felice, V.~Duvvuri, D.~A.~Easson, M.~Trodden 
and M.~S.~Turner, Phys.\ Rev.\ D {\bf 71}, 063513 (2005).  

\bibitem{allemandi}  G.~Allemandi, A.~Borowiec and M.~Francaviglia,  
Phys.\ Rev.\ D {\bf 70}, 103503 (2004).  

\bibitem{meng}  X.~Meng and P.~Wang,  
Class.\ Quant.\ Grav.\  {\bf 21}, 951 (2004). 

\bibitem{nojiri3}  S.~Nojiri and S.~D.~Odintsov,  
Phys.\ Rev.\ D {\bf 68}, 123512 (2003).



\bibitem{cappo1} S.~Capozziello, Int.\ J.\ Mod.\ Phys.\ D {\bf 11}, 
483 (2002).  



\bibitem{nojiri2}  S.~Nojiri and S.~D.~Odintsov,  
Phys.\ Lett.\ B {\bf 576}, 5 (2003).  

\bibitem{nojiri}  S.~Nojiri, arXiv:hep-th/0407099.  

\bibitem{dolgov}  A.~D.~Dolgov and M.~Kawasaki,  
Phys.\ Lett.\ B {\bf 573}, 1 (2003).  

\bibitem{cappo3} S.~Capozziello, V.~F.~Cardone and A.~Troisi,  
Phys.\ Rev.\ D {\bf 71}, 043503 (2005).

\bibitem{confprobs}  T.~Chiba,  
Phys.\ Lett.\ B {\bf 575}, 1 (2003).
 E.~E.~Flanagan, Class.\ Quant.\ Grav.\  {\bf 21}, 417 (2003);
Class.\ Quant.\ Grav.\  {\bf 21}, 3817 (2004);
G.~Magnano and L.~M.~Sokolowski,
Phys.\ Rev.\ D {\bf 50}, 5039 (1994).




\end{thebibliography}
\end{document}